\documentclass[conference,11pt,onecolumn,draftclsnofoot]{IEEEtran}
\usepackage{epstopdf}
\usepackage{algorithm,algorithmic}
\usepackage{graphicx}

\ifCLASSINFOpdf
\else
\fi

\begin{document}
%
\title{Online Rainbow Coloring In Graphs }

\author{\IEEEauthorblockN{Debasis Dwibedy, Rakesh Mohanty and Arun Khamari}
\IEEEauthorblockA{Department of Computer Science and Engineering\\
Veer Surendra Sai University of Technology\\
Burla, Odisha, India, 76018\\
Email: debasis.dwibedy@gmail.com, rakesh.iitmphd@gmail.com, arunkhamari11@gmail.com }}
\maketitle

\begin{abstract} 
Rainbow coloring is a special case of edge coloring, where there must be atleast one path between every distinct pair of vertices that consists of different color edges. Here, we may use the same color for the adjacent edges of a graph representing two different paths from a single vertex. In \textit{online rainbow coloring}, we have no priori knowledge about the vertices and edges  of the graph, infact the edges are available one by one. We have to color an edge as soon as it arrives and before the arrival of the next edge. We can not revoke the colorinng decision once it is made. According to our knowledge, there is no study of online rainbow coloring for graphs. In this paper, we make a first attempt to propose an online algorithm named \textit{Least Recently Used Color(LRUC)} for \textit{ online rainbow coloring}. We analyze the performance of \textit{LRUC} through competitive analysis. 
We show that \textit{LRUC} is optimal for line graph, tree and star graph. For  $1$-cyclic graph, \textit{LRUC} is shown to be ($2-\frac{2}{n}$)-competitive, where $n \geq 4$. We obtain the competitive ratios of $\frac{n-1}{3}$ and $n-1$ for wheel and complete graphs respectively, where $n$ is the number of vertices.

\end{abstract}


%
\IEEEpeerreviewmaketitle

\section{Introduction}

\subsection{Online Algorithm and Competitive analysis } \label{subsec: Online Algorithm and Competitive analysis} A computational problem is \textit{online} when the inputs of the problem are available one by one and an immediate action is desired after arrival of each new input. The algorithm designed for an online computational problem is called \textit{online algorithm}(\textit{ONL}) [1]. Here, the algorithm takes a sequence of decisions by considering present and past inputs without knowledge of future inputs. Formally, suppose we have an input sequence I=$<i_1,i_2,..,i_n>$ of finite size $n$, where $i_1$ is available at time $t=1$ , $i_2$ is available at $t=2$ and so on. At any given $t$ the input instances $I_t$ are only known and the input instances $I_{t'}$ are unknown to the algorithm, where $t'>t$. Optimal offline algorithm(\textit{OPTL}) is the one that has prior knowledge about the entire inputs and incurs minimum cost among all offline algorithm.\\
The performance of \textit{ONL} can be measured through \textit{competitive analysis} [2]. Here, the cost of \textit{ONL} is compared to the cost of the \textit{OPTL}. Let $ONL(I)$ and $OPTL(I)$ be the cost obtained by  $ONL$ and $OPTL$ respectively for processing of the input sequence $I$. $ONL$ can be $s$-competitive  if and only if  $ONL(I) \leq s. OPTL(I)+b$, where $b$ is a positive constant.
\subsubsection {Rainbow coloring} 
\label{subsec:Rainbow coloring }  
Rainbow coloring [3] of a non-trivial connected simple graph $G(v,e)$ is a special case of edge coloring where all vertices pair in $G(v,e)$ must have atleast $1$ \textit{rainbow path}. A rainbow path exists between a pair of vertices if and only if all edges in the path must have different colors. Formally, we define the rainbow coloring of the edges in $G(v,e)$ as $c: e(G) \rightarrow \{1,2,......k\}$, where $k \geq 1$. Here, $c$ is a rainbow $k$-coloring as $k$ different colors are used. Our objective is to minimize $k$ while making $G(v,e)$ rainbow colored. The minimum value of $k$ for $G(v,e)$ is the \textit{rainbow connection number} $rc(G)$ of $G(v,e)$.  
\subsubsection{Offline rainbow coloring} \label{subsec:Offline rainbow coloring }
 The rainbow coloring is offline when the algorithm has complete knowledge about  all vertices and edges of the graph prior to make a coloring decision. All the components of the graph are processed and colored simultaneously. Formally, suppose we have a non-trivial connected graph $G(v,e)$ with $n$ vertices $v_i(i=1,2,.....n)$ and $m$ edges $e_j(j=1,2,.....m)$. All $v_i$ and  $e_j$ are given as inputs to the offline algorithm in advance. The algorithm processes the input graph as a whole and produces a rainbow coloring of the edges of the graph. We denote $rc(G)$ obtains by \textit{OPTL} as $rc_{OPTL}(G)$.  For a simple illustration, we take a cyclic graph with $3$ vertices($C_3$) in figure $1(a)$ and present an offline rainbow coloring of $C_3$. 

\subsubsection{Online Rainbow Coloring} \label{subsec:Online Rainbow Coloring } In online rainbow coloring, the inputs are a non-trivial un-directed connected simple graph $G$ and a set of colors $c$. The output is a rainbow colored graph, where there must be atleast one \textit{rainbow path} between every distinct pair of vertices. Our goal is to use minimum colors while making $G$ rainbow colored. We have constraints such as the edges of $G$ are unknown at the begining and are available one by one in an order. The edge needs to be colored as soon as it arrives and prior to the arrival of the next edge. The coloring decision is irrevocable. The assumption is that the partial graph formed after the addition of each new edge must be connected. For simple illustration, we take $C_3$ in figure $1(b)$ and present an online rainbow coloring of $C_3$.


\subsubsection{Practical motivation} \label{subsec: Practical motivation}
Rainbow coloring can be used as a mechanism for frequency distribution among different links of a cellular network [4, 5]. If it is required distinct communication channels between a pair of mobile stations to communicate, then rainbow coloring can be applied to minimize the number of unique channels in the whole network. 

\subsubsection{Research motivation} \label{subsec: Research motivation}

 Computing $rc(G)$ of a non-trivial connected graph $G(v,e)$ has been proved to be NP-Hard [5]. It is non-trivial to decide whether a given coloring in an edge colored graph also holds the minimum colors to make the graph rainbow colored [5]. The problem becomes trivial when fixed number of colors are used. However, the problem is NP-Complete if the coloring is arbitrary [6].               

 \section{Background and Preliminaries}\label{sec:Background and Preliminaries}

\subsubsection{Definitions and Notations } \label{Definitions and Notations }
\begin{itemize}
	
\item A graph is \textit{Connected}  if it has atleast one path between each vertices pair [7].
\item \textit{Non-trivial graph($G$)} is a simple connected graph with atleast two edges [7].
\item \textit{Size of graph($m$)} denotes the total number of edges of the graph [7].

\item \textit{Degree of a vertex($deg(v)$)} denotes the number of edges incident on $v$ [7].
\item We call a vertex $v_i$ as \textit{pendant}, if $deg(v_i)=1$ [8]. 

\item \textit{Diameter($diam(G)$)} of a graph is the largest distance between two vertices $v_i, v_k$, which is maximum over the distances between all pairs of vertices in a graph [7, 8].



\item A graph $G(v,e)$ is \textit{rainbow connected } if there exists atleast one rainbow path between every pair of vertices [3].
\item \textit{Online Rainbow Connection Number($rc_{online}(G)$)} denotes the number of colors used through online rainbow coloring [3].



\item A simple connected non-trivial graph is \textit{Complete ($K_n$)} if all distinct pair of vertices are adjacent to each other [4, 7].

\item {Line graph($L_n$)} is a simple graph where the vertices are in one-one correspondence with the edges. Here, $v_{i}$ is adjacent to $v_{i-1}$ and $v_{i+1}$, for each $i$, $2\leq i \leq n-1$, where, $v_1$ and $v_n$ are adjacent to only $v_2$ and $v_{n-1}$ [4, 7]. 

\item In a \textit{Regular graph($R_n$)} with $n$ vertices, every vertex has equal degree. Suppose in $R_n$,  every vertices have degree equals to $r$, then $R_n$ is called $r$-regular [4, 7]. 

\item A {Cyclic graph($C_n$)} is basically a $2$-regular graph where number of vertices is equal to number of edges [7, 8]. If there exists only $1$ cycle in $C_n$, then we call $C_n$ as \textit{$1$-Cyclic}.

\item {Wheel graph($W_n$)} is a cyclic graph with an additional vertex which is connected to every other vertices of the cyclic graph $C_{n-1}$ [7].
\end{itemize}

%

\subsubsection{Overview of Related Work } \label{Related Work }
Offline rainbow coloring has been studied for graphs with limitless variants since the seminal work of Chartrand and et.al. in [3]. To acquaint with the state of the art literature and recent advancements in offline rainbow coloring, see survey [7]. In our concern, there is no study in the literature for online rainbow coloring in graphs. However, for the competitive analysis of our proposed online algorithm \textit{LRUC}, we must know about the optimal offline strategy for rainbow coloring in various graphs. Therefore, we present an overview of some important contributions and results in offline rainbow connections of graphs as follows.\\
Chartrand and et.al. [3] introduced \textit{rainbow coloring} in graphs. They showed that $rc(K_n)=1$ and $rc(T_n)=m$. They proved  $rc(W_n)=3$, for $n \geq 8$ and $rc(C_n)=\lceil\frac{n}{2}\rceil$, where $n \geq 4$. For the complete bipertite graph $K_{p,q}$, they obtained $rc(K_{p,q})=2$. Caro and et. al. [9] studied the rainbow connections in graph ($G$) with minimum degree of $G$. They obtained $rc(G)< \frac{5}{6}n$ for $\delta(G) \geq 3$, where $\delta(G)$ is the minimum degree of $G$. They proved that $rc(G) \leq min \{n\frac{ln(\delta)}{\delta}(1+o_\delta)(1), n\frac{4ln(\delta)+3}{\delta}\}$ for connected graph $G$ with $n$ vertices and minimum degree $\delta(G)$. The hardness of rainbow coloring in graphs was studied in [4,5]. In [4], authors proved that computing $rc(G)$ for any $G$ is NP-Hard. In [5], it was shown that obtaining $rc(G) \leq k$ for any given $k$ is NP-Complete. Chartrand and et. al. [10] defined $k$-connectivity of $G$ as $rc_k(G)$. They obtained $rc_k(K_n)=2$, for any integer $k \geq 2$ if there exists an integer $f(k)$, where  $f(k) \leq n$. In bipartite graph, they showed that for every $k \geq 2$, there is an integer $r$ such that $rc_K(K_{r,r})=3$. Krivelevich and Yuster [11] defined rainbow vertex connection $rvc(G)$ for any $G$. They proved that $rvc(G) < \frac{11n}{\delta}$. Schiermeyer [12] addressed the conjecture of Caro and et. al. [9]. He proved that $rc(G) < \frac{3n}{4}$ for $\delta \geq 3$. Chandran and et. al. [13] showed that $diam(G) \leq rc(G) \leq diam(G)+1$, where $G$ is an interval graph and $\delta \geq 2$. They proved $rc(G)=diam(G)$ if $G$ is an unit interval graph. For circular arc graph, they obtained the inequality $diam(G) \leq rc(G) \leq diam(G)+4$. Chartrand and et. al. [14] defined $k$-rainbow coloring as an edge coloring of $G$ such that for every set $A \subseteq k$ vertices of $G$, there exists a rainbow tree $T_n$ in $G$ such that $A \subseteq v(T)$, where $v(T)$ is the number of vertices of $T_n$. They defined $k$-rainbow index $rx_k(G)$ as the minimum number of colors required to $k$-rainbow color $G$. They showed that $rx_k(G)=n-2$ if $k=3$ and girth $g \geq4$. For uni cyclic graph of order $n \geq 3$, they obtained $rx_k(G)=n-1$. Li and Sun [15] addressed the open question put by Chartrand et. al. in [10] to determine $rc_k(K_{r,r})$. They showed that $rc_k(K_{r,r})=3$, where $r \geq 2k \lceil \frac{k}{2}\rceil$ and $k \geq 2$. In [16], Li and Sun studied for computing $rc(G)$, where $G$ is a line graph($L_n$) that consists of triangles. They obtained two upper bounds on $rc(G)$ for $L_n$ in terms of number of edge disjoint triangles of $L_n$. Li and et. al. [17] obtained $rc(G) \leq 5$ if $G$ is a bridge less graph and $diam(G)=2$. They showed that $rc(G) \leq k+2$ for any connected $G$ with $diam(G)=2$ and $k$ bridges, where $k \geq 1$. Li and et. al. [18] proved that $rc(G) \leq \lceil \frac{n}{2}\rceil$ for $n \geq 3$, where $G$ is a $2$-connected graph. Dudek and et. al. [19] studied rainbow connection of random $r$-regular graph $G(n,r)$ of order $n$, where $r \geq 4$. They proved that $rc(G)= O(logn)$.

\section{Our Contribution and Results } \label{Our Contribution and Results }
 \subsubsection{Online Rainbow Coloring Algorithm} According to our knowledge their is no study of online algorithm for rainbow coloring in the literature. We make a first attempt to propose an online rainbow coloring algorithm named \textit{Least Recently Used Color}(\textit{LRUC}) for various types of graphs such as line, tree, star, cyclic, wheel, complete and bipartite. The pseudocode of \textit{LRUC} algorithm is presented as follows. 
 
   \label{Online Algorithm}
  \begin{algorithm}
  	\caption{LRUC}
  	\begin{algorithmic}[1]
  		\scriptsize
  		\STATE Initially, i=1, j=1, Set of Colors $c:\{c_1\}$ \\
  		\STATE Assign color $c_1$ to the first edge $e_1$.\\
  		\STATE j=j+1.
  		\STATE WHILE a new edge $e_j$ arrives\\
  		\STATE \hspace*{0.2cm} BEGIN\\
  		\STATE \hspace*{0.4cm} IF $e_j$ is adjacent to only one already arrived edge.\\
  		\STATE \hspace*{0.6cm} THEN i=i+1.\\
  		\STATE \hspace*{0.6cm} Assign a new color $c_i$ to $e_j$.\\
  		\STATE \hspace*{0.6cm} Insert the new color $c_i$ to the set of colors $c$. \\
  		\STATE \hspace*{0.4cm} END IF. \\
  		\STATE \hspace*{0.4cm} ELSE IF $e_j$  is adjacent to atleast two already arrived edges.\\
  		\STATE \hspace*{0.6cm} IF one of the vertex of $e_j$ has degree $1$.\\
  		\STATE \hspace*{0.8cm} THEN i=i+1.\\
  		\STATE \hspace*{0.8cm} Assign a new color $c_i$ to $e_j$.\\
  		\STATE \hspace*{0.8cm} Insert $c_i$ to the set of colors $c$.\\
  		\STATE \hspace*{0.6cm} END IF.\\
  		\STATE \hspace*{0.6cm} ELSE IF both the vertices of $e_j$ has degree atleast $2$.
  		\STATE \hspace*{0.8cm} THEN Assign the least recently used color from the set of colors.\\
  		\STATE \hspace*{0.6cm} END ELSE IF. \\
  		\STATE \hspace*{0.4cm} END ELSE IF.\\
  		\STATE \hspace*{0.4cm} j=j+1\\
  		\STATE \hspace*{0.2cm} END WHILE\\
  		\STATE Return Set of Colors $c$. \\
 		\STATE END
 		\end{algorithmic}
 	\end{algorithm}

%
%
%


 \textbf{Theorem. 1. LRUC is $1$-competitive for class $A$, where $A \in G$ and  $A=\{L_n, T_n, S_n\}$}.\\
\textit{Proof:} For the competitive analysis of \textit{LRUC}, we have to first compute the cost of \textit{OPTL}, then the cost of \textit{LRUC}. The ratio between the cost of \textit{LRUC} and  cost of \textit{OPTL} gives us the competitive ratio of LRUC for any $G$, which is the performance indicator for \textit{LRUC}. So, first we verify for ($L_n$) as follows.\\
\textit{ Computation of OPTL}: In $L_n$, all edges are adjacent to its previous and next edge. However, the first edge $e_1$ is adjacent to only its next edge $e_2$ and the last edge $e_{n-1}$ is adjacent to only its previous edge $e_{n-2}$. This structure of a line graph sets up only one path between the extreme vertices pair($v_1, v_n$) through the edges $e_1, e_2....e_{n-1}$. So, to satisfy the rainbow coloring property, \textit{OPTL} uses distinct colors for all edges from $e_1$ to $e_{n-1}$.\\ Therefore, $rc_{OPTL}(L_n) \leq n-1$. \hspace*{10.8cm}(1)\\   
\textit{Computation of LRUC}: \textit{LRUC} assigns a new color to each incoming edge $e_j$ of the line graph($L_n$). Because each $e_j$, where $1 \leq j \leq n-1$ has atleast one \textit{pendant} vertex. Therefore, the minimum number of colors required for $L_n$  is equal to $n-1$. Formally suppose we have $n$ vertices, then the geodesic from $v_1$ to $v_n$ contain $n-1$ edges where all edges must be colored different.\\ Therefore, we have $rc_{LRUC}(L_n) \leq n-1$. \hspace*{9.4cm} (2)\\
From equations (1) and (2), we can have: $\frac{rc_{LRUC}(L_n)}{rc_{OPTL}(L_n)} = \frac{n-1}{n-1}=1$.\\
Now, we verify for ($T_n$) as follows.\\
\textit{Computation of OPTL}: \textit{OPTL} uses $m$ colors to make a tree rainbow colored, where $m$ is the size of the tree [3]. \\ Therefore, we have $rc_{OPTL}(T_n)= m$. \hspace*{9.8cm} (3) \\
\textit{Computation of LRUC}: We can use at most $m$ colors to make any graph rainbow colored. Suppose we use $k$ colors for rainbow coloring in tree, where $k < m$. The edges of the tree are available one by one, so, every newly arrive edge must have a \textit{pendant} vertex. Hence, the new edge becomes the part of the unique path from the \textit{pendant} vertex to every other known vertices of the tree. Therefore, we have to use different colors to each incoming edge. If we use any existing color, then there must be an altenate path from the \textit{pendant} vertex to atleast one known vertex. This indicates the existance of a cycle in the tree, which is a contradiction. So, we must use the number of colors equal to the size of the tree not less than that. Therefore, we have $rc_{LRUC}(T_n)=m$. \hspace*{10.5cm} (4)  \\
From equations (3) and (4), we have $\frac{rc_{LRUC}(T_n)}{rc_{OPTL}(T_n)} = \frac{m}{m}=1$. \\
Now, we verify for ($S_n$) as follows.\\
\textit{Computation of OPTL:} In $S_n$, all edges are adjacent to each other through a central vertex. Let the central vertex be $v_1$. So, every distinct ($v_i, v_k$) pairs, where $2 \leq (i, k) \leq n$ have an unique path of length $2$ and for each $v_i$, where $2 \leq i \leq n$, the path between ($v_1, v_i$) is of length $1$. Therefore, $S_n$ consists of $n-1$ \textit{pendant} vertices. \textit{OPTL} initiates the rainbow coloring by choosing any of the \textit{pendant} vertex (let, $v_2$) and explore the paths to all other $v_i$, where, $3 \leq i \leq n$. So, \textit{OPTL} uses  $2$ colors that satisfies the rainbow coloring property between $v_2$ and every other $v_i$. Subsequently, the next \textit{pendant} vertex(let, $v_3$) is chosen and the unknown paths to other $v_i$, where, $4 \leq i \leq n$ are explored. Here, we use an additional color besides the $2$ earlier used colors because if we use an existing color, then there would be no \textit{rainbow path} either between $v_2$ and atleast one of the other $v_i$, where $3 \leq i \leq n$ or between $v_3$ and atleast one of the other $v_i$, where $2 \leq i \leq n$ (as $S_n$ is a simple graph, we can ignore edges from $v_i$ to $v_i$). Similarly, by considering rest of the $n-3$ \textit{pendant} vertices one by one, we use an additional $n-3$ colors. So, in total we use $n-1$ colors, which is equal to the size of $S_n$ i.e. $m$.\\ Therefore, we have $rc_{OPTL}(S_n)= m$. \hspace*{9.5cm} (5) \\
\textit{Computation of LRUC}: Online Rainbow coloring of $T_n$ and $S_n$ are identical. We refer to the computation of \textit{LRUC} for $T_n$ as a proof for online rainbow coloring of $S_n$.\\ Therefore, we have $rc_{LRUC}(S_n)= m$. \hspace* {9.5cm}  (6) \\
From equations (5) and (6) we have $\frac{rc_{LRUC}(S_n)}{rc_{OPTL}(S_n)} = \frac{m}{m}=1$.  \\     
  
\textbf{Theorem. 2. LRUC is ($2-\frac{2}{n}$)-competitive for $C_n$, where $C_n$ is $1$-Cyclic and $n \geq 4$}.\\
  
\textit{Proof:} For the optimal offline rainbow coloring, we refer to the policy of Chartrand and et.al. [3]. OPTL uses $\lceil\frac{n}{2}\rceil$ colors for $C_n$, where $n \geq 4$ [3].\\
Therefore, we have $rc_{OPTL}(C_n)=\lceil\frac{n}{2}\rceil $ \hspace* {9.4cm} (7)

\textit{Computation of LRUC}: Suppose $C_n$ contains $1$ cycle of length $n$ with vertices ($v_1, v_2,....v_{n-1}, v_n, v_1$). In worst case, edges($e_j$) arrive one by one in the order: $e_1(v_1, v_2), e_2(v_2, v_3)....e_{n-1}(v_{n-1}, v_n)$. \textit{LRUC} assigns a new color to each $e_j$ till the arrival of $e_{n-1}$. In any incoming $e_j$, where $2 \leq j \leq n-1$, if we use an existing color, then there would not be a \textit{rainbow path} between extreme pair of vertices $v_1$ and $v_{n+1}$ as both are \textit{pendant} now. So, we bound to use $n-1$ colors. The arrival of $e_n(v_n, v_1)$ completes the cycle and can be colored with $c_1$ as $v_1, v_n$ are adjacent now.\\
Therefore, we have $rc_{LRUC}(C_n)= n-1$. \hspace* {9.3cm} (8)     

From equations (7) and (8), we have: $\frac{rc_{LRUC}(C_n)}{rc_{OPTL}(C_n)} = \frac{n-1}{\lceil\frac{n}{2}\rceil} \leq \frac{n-1}{(\frac{n}{2})} \leq \frac{2(n-1)}{n} \leq \frac{2n-2}{n} \leq 2-\frac{2}{n}$. \\

  
\textbf{Theorem. 3. LRUC is ($\frac{n-1}{3}$)-competitive for $W_n$, where $n \geq 8$}.\\\\
\textit{Proof}: \textit{OPTL} uses $3$ colors for rainbow coloring in $W_n$, where $n \geq 8$ [3].\\
Therefore, $rc_{OPTL}(W_n)= 3$. \hspace* {11.1cm} (9) \\
\textit{Computation of LRUC}: In $W_n$, let $v_1$ be the central vertex connected to each $v_i$, where $i \leq 2 \leq n$. In the worst case, first the internal $n-1$ edges arrive one by one in any order. The internal $n-1$ edges form $S_n$ by connecting $v_1$ to each $v_i$, where $i \leq 2 \leq n$. So, \textit{LRUC} uses $n-1$ colors to the internal $n-1$ edges due to equation (6). Irrespective of the order of arrival of the external $n-1$ edges connecting distinct ($v_i, v_k$), where $2 \leq (i, k) \leq n$, \textit{LRUC} uses least recently used color to each $e_j$, where $n \leq j \leq 2n-2$.\\
Therefore, $rc_{LRUC}(W_n)=n-1$. \hspace* {10.5cm} (10)\\
From equations (9) and (10), we have $\frac{rc_{LRUC}(W_n)}{rc_{OPTL}(W_n)}=\frac{n-1}{3}$. \\

  
\textbf{Theorem. 4. LRUC is ($n-1$)-competitive for $K_n$}. \\\\
\textit{Proof}: In $K_n$, all distinct pair of vertices($v_i, v_k$) are adjacent to each other, where $1 \leq (i,k) \leq n$. So, \textit{OPTL} assigns $1$ color to each $e_j$ [3].\\
Therefore, $rc_{OPTL}(K_n)=1$ \hspace* {11.0cm} (11)\\
\textit{Computation of LRUC}: In the worst case, the first $n-1$ edges of $K_n$ arrive in the order of the \textit{line graph}($L_n$) i.e. $e_1(v_1, v_2), e_2(v_2, v_3)...e_{n-1}(v_{n-1}, v_n)$. So, \textit{LRUC} uses $n-1$ different colors to the first $n-1$ edges due to equation (2). For rest $\frac{n^2-3n+2}{2}$ edges, \textit{LRUC} uses least recently used color at each time when $e_j$ arrives irrespective of its order of arrival because the vertices of each $e_j$   have degree atleast $2$ as each $e_j$ must adjacent to atleast $2$ already arrived edges, where $n \leq j \leq \frac{n(n-1)}{2}$.\\
Therefore, $rc_{LRUC}(K_n)=n-1$ \hspace* {10.2cm} (12)\\
From equations (11) and (12) we have $\frac{rc_{LRUC}(K_n)}{rc_{OPTL}(K_n)}=\frac{n-1}{1}=n-1$. \\
  

  

  \section{Conclusion } \label{Conclusion}
  
 We have studied rainbow coloring of graphs in an algorithmic perspective, where the components of the graph such as edges available one by one to the algorithm unlike the whole graph in traditional rainbow coloring problem of Chartrand and et. al. [3].  We have proposed the first online algorithm named \textit{LRUC} for the online rainbow coloring problem. We have proved \textit{LRUC} be the \textit{optimal} online algorithm for special classes of graphs such as line, tree and star. \textit{LRUC} has been shown to be (2-$\frac{2}{n}$)-competitive for cyclic graph which contains only one cycle and atleast four vertices. We have obtained the competitive ratios of $\frac{n-1}{3}$ and $n-1$ for wheel and complete graph respectively. We have observed that the performance of \textit{LRUC} depends on the order of availability of the edges of any graph. Due to lack of knowledge about the whole graph, \textit{LRUC} obtains greater rainbow connection number than the optimal offline algorithm(\textit{OPTL}). Therefore, it will be interesting to investigate for a better online algorithm that perform near equal to \textit{OPTL}. Further, \textit{LRUC} can be studied for online rainbow coloring in other graphs.

\end{document}